\documentclass[11pt]{article}
\usepackage{amsmath,epsf,amssymb,latexsym,enumerate,cite,shadow,array,color}
\usepackage[ps,dvips,matrix,arrow,frame,import,curve,color]{xy}
\DeclareFontFamily{U}{rsf}{}
\DeclareFontShape{U}{rsf}{m}{n}{
  <5> <6> rsfs5 <7> <8> <9> rsfs7 <10-> rsfs10}{}
\DeclareMathAlphabet\Scr{U}{rsf}{m}{n}

\makeatletter
\@addtoreset{equation}{section}
\makeatother

\def\be{\begin{equation}}
\def\ee{\end{equation}}

\def\ba{\begin{array}}
\def\ea{\end{array}}

\newcommand{\bea}{\begin{eqnarray}}
\newcommand{\eea}{\end{eqnarray}}

\begin{document}

\begin{titlepage}
\begin{flushright}
SU-ITP-2011/31\\
\today
\end{flushright}
\vskip  2 cm

\vspace{24pt}

\begin{center}
{ \LARGE \textbf{  N=8   Counterterms and  $ E_{7(7)}$  Current Conservation }}

\vspace{28pt}

\vspace{28pt}

{\bf Renata Kallosh}

\

 \textsl{Department of Physics \emph{and} Stanford Institute for Theoretical Physics, \\
 Stanford University
 Stanford, CA 94305-4060, USA }

\vspace{10pt}

\vspace{10pt}

\vspace{24pt}

\

\end{center}

\begin{abstract}

We examine conservation of the $E_{7(7)}$ Noether-Gaillard-Zumino current in the presence of N=8 supergravity counterterms using the momentum space helicity formalism, which significantly simplifies the calculations. The main result is that the 4-point counterterms at any loop order $L$ are forbidden by the $E_{7(7)}$ current conservation  identity. We also clarify the relation between linearized and full non-linear superinvariants as candidate counterterms. This enables us  to show that all $n$-point counterterms at $L=7, 8$ are forbidden since they provide a non-linear completions of the 4-point ones.  This supports and exemplifies our  general proof in arXiv:1103.4115 of perturbative UV finiteness of N=8 supergravity.

\end{abstract}

\end{titlepage}

\tableofcontents

\newpage
\section{Introduction}

It has been shown in \cite{Kallosh:2010kk,Kallosh:2011dp} that N=8 \cite{Cremmer:1979up}  perturbative supergravity is UV finite.\footnote{We refer the  reader to a  recent discussion of perturbative and non-perturbative aspects of N=8 supergravity in \cite{Ferrara:2011er}.}
 The argument in \cite{Kallosh:2010kk} is based on the properties of the light-cone superspace and helicity formalism method.  The claim is that the candidate counterterms\footnote{At present on shell light-cone counterterms have not been constructed  in the real light-cone superspace \cite{Brink}.}
in the real and chiral light-cone superspaces are incompatible and therefore do not support the UV divergences, if the theory is anomaly-free. In particular, this argument relies on the equivalence between light-cone and Lorentz covariant computations in perturbative N=8 supergravity.

In \cite{Kallosh:2011dp} a more recent analysis of UV divergences was performed in the Lorentz covariant setting.  Two related proofs were proposed.

 The first Lorentz covariant proof in \cite{Kallosh:2011dp}  is based on an observation
of  the uniqueness of the Lorentz and $SU(8)$ covariant, $E_{7(7)}$ invariant unitarity constraint expressing the 56-dimensional  $E_{7(7)}$ doublet via 28 independent vectors, in agreement with the ${E_{7(7)}\over SU(8)}$ coset space geometry. It was shown in
\cite{Bossard:2010dq} that in N=8 supergravity the non-linear  $E_{7(7)}$ symmetry  is non-anomalous to all orders, which supports the unitarity argument in  \cite{Kallosh:2011dp}.

 The second Lorentz covariant proof  is based on  a suggestion  in \cite{Kallosh:2011dp} that the $E_{7(7)}$ symmetry of N=8 supergravity  has to be viewed in the context of continuos global symmetry which requires the Noether-Gaillard-Zumino  current conservation \cite{Gaillard:1981rj}, \cite{Kallosh:2008ic}. This requirement is necessary for the $E_{7(7)}$ symmetry of a complete theory; it is significantly stronger than the previously used condition that the counterterms have to be invariant under $E_{7(7)}$ symmetry. The non-linear candidate counterterms, invariant under classical $E_{7(7)}$ symmetry have been constructed long time ago in \cite{Howe:1980th,{Kallosh:1980fi}}, using the on shell covariant superspace \cite{Brink:1979nt}. Their existence was  the basic reason to abandon perturbative N=8 supergravity. However,  the $E_{7(7)}$ current conservation, deformed by candidate counterterms, was not studied in N=8 supergravity until recently. In \cite{Kallosh:2011dp} we studied $E_{7(7)}$ current conservation in the coordinate space, in this paper we study it in momentum space, using helicity formalism.

Meanwhile the 3-loop UV divergence in N=8 supergravity, supported by the candidate linearized counterterm in \cite{Kallosh:1980fi}, was shown to be absent by explicit computations in \cite{Bern:2007hh}. One of the explanation is due to $E_{7(7)}$ symmetry \cite{Brodel:2009hu}. The argument in \cite{Brodel:2009hu} about $E_{7(7)}$ forbidding the 3-loop UV divergence in N=8 supergravity  was specific for the 3-loop case only.  However, the candidate counterterms  have been constructed in higher loops not only in superspace  \cite{Howe:1980th,Kallosh:1980fi} but also using the helicity amplitude methods at the linearized level in \cite{Kallosh:2009db}, \cite{Kallosh:2010kk} and in full generality in \cite{Elvang:2010jv}.

The $E_{7(7)}$ current conservation studied in  \cite{Kallosh:2011dp} explains the absence of the 3-loop divergence discovered in \cite{Bern:2007hh}, but it also applies to higher loops. The proposal of \cite{Kallosh:2011dp} is that one has to check if the N=8 supergravity action, deformed by counterterm actions,
\be
S= S_{cl} + S_{CT} \ ,
\ee
still has  $E_{7(7)}$ symmetry.  Namely, given the action which depends on 28 vectors $F_{\mu\nu}^{IJ}$, one defines the dual field strength as a derivative of the deformed action
\be
\tilde G^{\mu\nu IJ }\equiv 2 {\delta S\over \delta  F_{\mu\nu}^{IJ}}\, .
\label{G}
\ee
The  56-component $E_{7(7)}$ doublet 
$(F^{IJ}, \; G_{IJ})$
consists of 28 original vectors fields $F=d{\cal A}$ and 28 dual vector fields $G=d{\cal B}$. The dual field strength is a functional of the original ones and scalars,
\be
G= G(F, \phi)\, .
\ee
This relation, which we call unitarity constraint and describe in detail in eqs. (5.4) and (5.8) in \cite{Kallosh:2011dp}, does not admit deformations consistent with global $E_{7(7)}$ and local Lorentz and $SU(8)$ symmetries. This relation provides the first general Lorentz covariant proof of perturbative finiteness. 

The second Lorentz covariant proof in \cite{Kallosh:2011dp}, related to the first one,  is based on $E_{7(7)}$ current conservation identity
\be \boxed{
\langle\;  \alpha \; | \partial_\mu  J^\mu | \; \beta \; \rangle =0}
\label{NGZ}\ee
Here $\langle \alpha |$ and $| \beta \rangle$ are physical states of the theory. $ J^\mu \in E_{7(7)}$ is the  Noether-Gaillard-Zumino current \cite{Gaillard:1981rj}, \cite{ Kallosh:2008ic}, \cite{Bossard:2010dq}. It has  133-components  in the fundamental 56-dimensional representation given by  \begin{eqnarray}\label{gauge}
   G_{E_{7(7)}}=\left(
                                        \begin{array}{cc}
                                         \Lambda_{IJ}{}^{KL} & \Sigma _{IJPQ} \\
                                          \Sigma^{MNKL} & \Lambda ^{MN}{}_{PQ} \\
                                        \end{array}
                                      \right)\ .
\label{fund}\end{eqnarray}
The non-trivial part comes from the 70-component vector dependent part of the Noether-Gaillard-Zumino current $ J^\mu_B\equiv   J^\mu_{IJKL} B^{IJKL}$ sandwiched between physical states containing  vectors. In such case
\be
\langle \alpha_v | \partial_\mu  J^\mu_B | \beta_v \rangle =  \langle \alpha_v | \tilde G^{\mu\nu}_{IJ}\, B^{IJKL} \, G_{\mu\nu KL} | \beta_v \rangle =0
\label{NGZ1}\ee
Here $\langle \alpha_v |$ and $| \beta_v \rangle$ are physical states containing  vectors. The
 70 real parameters $B$  are related to the one in the (\ref{fund}) as follows \cite{Kallosh:2008ic}
\be
B= \mbox{Im}\; \Lambda+\mbox{Im}\; \Sigma
\ee
Their role in $E_{7(7)}$ duality is to mix Bianchi identities $\partial_{\mu} \tilde F^{\mu\nu IJ}=0$ with equations of motion $\partial_{\mu} \tilde G^{\mu\nu}_{ KL}=0 $.
\be
\Delta  \; \partial_{\mu} \tilde F^{\mu\nu IJ}= B^{IJKL} \; \partial_{\mu} \tilde G^{\mu\nu}_{ KL}
\ee
These are  off-diagonal terms in the $Sp(56)$ embedding of the $E_{7(7)}$  performed in \cite{Kallosh:2008ic} ,  and therefore they play non-trivial role in testing the effect of deformation of supergravity action by the counterterms.

In \cite{Kallosh:2011dp} the explicit computation of the expression in the left hand side  of (\ref{NGZ1}) was performed for the 3-loop counterterm \cite{Kallosh:1980fi,Freedman:2011uc} and it was shown to contradict the right hand side  of the identity, unless there is no 3-loop divergence.
Here we will first switch to momentum space and helicity formalism, which makes the explicit computation of L-loop contributions to $E_{7(7)}$ identity manageable. We will show how it works for the 3-loop case and proceed with the 4-point $L$-loop case. The insertion of functionals of Mandelstam variables for increasing dimension of the 4-point counterterms allows a complete classification of all possible independent 4-point $L$-loop counterterms, using the results in \cite{Green:1999pv}
for the  low-energy expansion of the  superstring amplitudes. This leads to a statement that all loop $L$, 4-point counterterms violate the $E_{7(7)}$ current conservation. It is based on the fact that the number of independent insertions of Mandelstam variables is discrete and finite whereas the identity has to be valid at continuous values of
Mandelstam variables.

To study the  $n>4$ case we compare the  non-linear as well as linear counterterms in superspace \cite{Howe:1980th,Kallosh:1980fi,Kallosh:2010kk} with the linear ones derived in  \cite{Elvang:2010jv}. We explain the difference between linearized and geometric superfields. In particular we have to use the fact that the linearized scalar superfield $W_{ijkl}(x, \theta)$ has dimension zero and any power of it, $W^n$, still has dimension zero. Therefore the linearized superfield actions, with manifest global supersymmetry, are easy to construct in superspace for the $n$-point amplitudes. The corresponding superinvariants are given in a symbolic form as  
\be
\kappa^{2(L-1)} (D_\alpha^m)^{2a} (D_{\dot \alpha r})^{2b} (D_{\beta \dot \beta})^c \Big (W_{ijkl}(x, \theta)\Big )^n
\ee 
and include all possible $SU(8)$ invariant combinations of scalar superfields and its supercovariant derivatives sprinkled over the superfields in an arbitrary way. For the $n$-point $L$-loop linearized superinvariants
\be
L= 7 +{a+b+c\over 2}
\ee
For example,  we explain the linearized superspace origin  of counterterms in \cite{Elvang:2010jv} which have N$^6$MHV 16-point contribution at $L=7$ and  N$^5$MHV 14-point contribution at $L=8$.

It is important, however, that the geometric superinvariants which have non-linear exact local supersymmetry can be constructed only using the geometric superfields, torsion and curvature. The smallest dimension torsion is a gaugini superfield $\chi_{\alpha \, ijk}$ which has dimension 1/2. At the 7 and 8 loop level only the superinvariants which start with the 4-point amplitudes are allowed by dimension. These are ruled out by $E_{7(7)}$ identity as we  show in the first part of the paper.  Therefore for $L=7,8$ we can give a detailed explanation why $E_{7(7)}$ current conservation forbids all $n$-point counterterms.

Thus, here we provide an explicit set of examples of the Lorentz covariant  proof of UV finiteness of N=8 supergravity
 \cite{Kallosh:2011dp}  explaining in details how the candidate counterterms  break $E_{7(7)}$ current conservation.

\section{$E_{7(7)}$ current conservation identity}

The $E_{7(7)}$ symmetry of the deformed action requires that  the deformed equations of motion  $  \partial_{\mu} \tilde G^{\mu\nu}=0$ transform into Bianchi identity $\partial_{\mu} \tilde F^{\mu\nu}=0$ as follows
\begin{eqnarray}\label{symplecticdual}
\Delta \left(
                      \begin{array}{cc}
                  \partial_{\mu} \tilde F^{\mu\nu} \\
                  \partial_{\mu} \tilde G^{\mu\nu} \\
                      \end{array}
                    \right)\ =\left(
                      \begin{array}{cc}
                      A&  B \\
                        C & D \\
                      \end{array}
                    \right)  \left(
                      \begin{array}{cc}
                  \partial_{\mu} \tilde F^{\mu\nu} \\
                  \partial_{\mu} \tilde G^{\mu\nu} \\
                      \end{array}
                    \right) \ .
\end{eqnarray}
It  requires the dual field strength
$ G(F, \phi)$ to transform according to $E_{7(7)}$ symmetry, so that
\begin{eqnarray}\label{symplecticdual}
\Delta \left(
                      \begin{array}{cc}
                   F \\
                  G \\
                      \end{array}
                    \right)\ =\left(
                      \begin{array}{cc}
                      A&  B \\
                        C & D \\
                      \end{array}
                    \right)  \left(
                      \begin{array}{cc}
                  F \\
                  G \\
                      \end{array}
                    \right) \ .
\end{eqnarray}
Here $A,B,C,D$ are real global infinitesimal parameters of $E_{7(7)}$ embedded into $Sp(56)$. Note that the vectors have a homogeneous transformation under $E_{7(7)}$, whereas the scalars start with the inhomogeneous term.
\be
\Delta\phi_{IJKL}= \Sigma_{IJKL}+...
\ee
where $B= \mbox{Im}\; \Sigma+...$.
Most of the  studies of  $E_{7(7)}$ symmetry of N=8 supergravity were associated with the  soft momentum limits of scalars. Here we take advantage of the action of $E_{7(7)}$ symmetry on vectors.

In presence of counterterms  $ G^{\mu\nu}= G^{\mu\nu}_0 + \hat G^{\mu\nu}$ consists of the classical part $ G_{0 }$ and deformation $\hat { G} $ caused by the deformation of the action by the counterterms $S_{CT}$.
  The corresponding Noether-Gaillard-Zumino \cite{Gaillard:1981rj} identity was given in  \cite{Kallosh:2011dp} in the form
\be
 {\delta \over \delta F^\Lambda }\int  d^4x \Big (\hat {\tilde G}(x) B \hat G(x)\Big ) =0\, ,   \qquad \rm{where} \qquad \hat {\tilde G}=2 {\delta S_{CT}\over \delta  F_{\mu\nu}}
\label{critical}\ee
and $B$ is an off-diagonal part of the $E_{7(7)}$ transformations in (\ref{symplecticdual}), mixing Bianchi identities with deformed equations of motion.

Note that once the equations of motion for deformed supergravity are solved, $ \partial_{\mu} \tilde G^{\mu\nu}=0$,  the dual potential is available, $G_{\mu\nu}= \partial _\mu {\cal B}_\nu- \partial _\nu {\cal B}_\mu$.
The vector part of the $B$-component of the NGZ current \cite{Gaillard:1981rj} is given by
\be
J^\mu_v (x) = {1\over 2} \tilde G^{\mu\nu} B \,{\cal B}_\nu\, ,   \qquad \partial _\mu  J^\mu_v (x) = \tilde G^{\mu\nu} B G_{\mu\nu}
\ee
The split of the current into the classical part and the deformed part corresponds to
\be
 \partial _\mu  J^\mu_v (x)=  \partial _\mu \hat J^\mu_{v 0} (x)+  \partial _\mu \hat J^\mu_v(x) = \tilde G^{\mu\nu} B G_{\mu\nu}=    \hat {\tilde G}^{\mu\nu} B \hat G_{\mu\nu}
\ee
since according to (\ref{critical}) the terms
\be
 \tilde G^{\mu\nu}_0 B G_{\mu\nu 0 } + 2 G^{\mu\nu}_0 B \hat {\tilde G}^{\mu\nu}
 \ee
do not contribute to the identity. The term $\tilde G^{\mu\nu}_0 B G_{\mu\nu 0 }$ is part of the classical $E_{7(7)}$ current conservation. The term $ G^{\mu\nu}_0 B  \hat {\tilde G}^{\mu\nu} = 2 G^{\mu\nu}_0 B   {\delta S_{CT}\over \delta  F_{\mu\nu}}$ vanishes for deformation due to
the counterterms invariant under classical $E_{7(7)}$ symmetry. This is quite different from Born-Infeld type non-linear dualities  \cite{Gaillard:1981rj} where higher order in $F$ terms are compensating order by order and where the corresponding term $ G^{\mu\nu}_0 B  \hat {\tilde G}^{\mu\nu}$ does not vanish.

\section{ 3-loop 4-point case in momentum space}
Here we re-derive the effect of the 3-loop counterterm on $E_{7(7)}$ identity presented in \cite{Kallosh:2011dp}. Instead of $x$-space we use momentum $p$-space here and helicity formalism. This simplifies the derivation and also allows to generalize it to higher loops.
We start by  writing the  gravity-vector part of the 3-loop counter term in momentum-space in a symbolic form without indices as
\be
S^{3-loop}_{ _{(\partial F)^2 R^2}} = x_3 \, \kappa^4 \int \prod_{i=1}^{4} d^4p_i \delta \Big (\sum p_i \Big ) R(p_1) \partial F(p_2) \bar R(p_3) \partial\bar F(p_4)\,.
\ee
Here $R, F$ etc. are  general  multi-spinor functions, not yet written in terms of the $\lambda_\alpha(p)$, and the $p_i$ are general 4-vectors, not yet null.    This can be considered as an off-shell extension of the term in \cite{Freedman:2011uc} in eq.
\be
S^{3loop}_{ _{(\partial F)^2 R^2}}
= x_3 \, \kappa^4 \int d^4x \,  R_{\dot \alpha \dot \beta \dot \gamma\dot \delta} \, F^{\dot \alpha \dot \beta ij } \, \partial^{\dot \gamma \gamma} \partial^{\dot \delta \delta} \, F^{\alpha \beta}_{ij} R_{ \alpha  \beta  \gamma \delta} \ .
\ee

To test the effect of the 3-loop counterterm on  $E_{7(7)}$ identity we need to compute the dual field strength defined by the variation of the action over $F$. We will now perform this variation in the momentum space. We  take ${\delta S^{3loop} \over \delta \bar F}$ to obtain
\be
\hat{G}(p) = x_3 \, \kappa^4  \int\prod_{i=1}^{3} d^4p_i \delta(p+ \sum_{i=1}^{3} p_i)R(p_1) \partial \partial F(p_2) \bar R(p_3).
\label{Dan}\ee
  Now we form the integral
\be
B^{IJKL} \int d^4 p\,  \hat G_{\dot \alpha \dot \beta IJ} (p)  \, \hat G^{\dot \alpha \dot \beta}_{ KL} (p)
=( x_3 \, \kappa^4)^2  B^{IJKL}    X_{IJKL}\ ,
\label{int}\ee
which corresponds to space-time expression
\be
 B^{IJKL} \int  d^4x  \, \hat G_{\dot \alpha \dot \beta IJ}(x) \,  \hat G^{\dot \alpha \dot \beta}_{ KL}(x)
 \ee
 which we need for the $E_{7(7)}$ current conservation in the form:
\be
{\delta \over F_{\alpha \beta MN }}  B^{IJKL} \int  d^4x \Big ( \hat G_{\dot \alpha \dot \beta IJ} \,  \hat G^{\dot \alpha \dot \beta}_{ KL} -h.c.\Big ) ={\delta \over F_{\alpha \beta MN }(y)} \int d^4 x  \; \partial_\mu J^\mu (x)=0\ .
\label{int1}\ee
  Now we put on shell the multi-spinor forms of $R, F, \bar R$, see \cite{Kallosh:2008mq} and \cite{Freedman:2011uc} for details
\be
F_{\alpha \beta IJ}(p) \approx F_{\alpha \beta ij}(p)\Rightarrow \lambda_{\alpha}(p) \lambda_{\beta}(p)  A_{IJ} (p)
\label{dictionaryF} \ee
with $A_{IJ}(p) \Rightarrow A_{IJ}(p) \delta(p^2)$ with $p$ a null 4-vector. Do the same for
$R$ and $\bar R$
\be
R_{\alpha \beta\gamma\delta}(p) \Rightarrow \lambda_{\alpha}(p) \lambda_{\beta}(p) \lambda_{\gamma}(p) \lambda_{\delta}(p) h(p) \ , \qquad \bar R_{\dot \alpha \dot \beta\dot \gamma\dot \delta}(p) \Rightarrow  \bar \lambda_{\dot \alpha}(p) \bar \lambda_{\dot\beta}(p) \bar \lambda_{\dot \gamma}(p) \bar \lambda_{\dot\delta}(p) \bar h(p) \ .
\label{dictionaryC}\ee
and assume that the sum $p_1+p_2+p_3$ is also a null vector.  The purpose of this restriction to $\delta$-functions of null $p_i$ is to work in a limit in which there are no off-shell corrections to the counter term.

 Finally we may perform  the integral over the 7 momenta $p$ and $ p_1, p_2, p_3$ for the first $G$ and  $p_4, p_5, p_6$ for the second $G$ in (\ref{int}) and get a non-zero result.  There is one $\delta$-function left which imposes the overall conservation constraint,    $\delta(\sum_{i=1}^6 p_i)$.
 We find
 using (\ref{dictionaryF}), (\ref{dictionaryC}) and $\partial^{\dot \gamma \gamma} \Rightarrow p^{\dot \gamma \gamma} \Rightarrow \bar \lambda^{\dot \gamma} \lambda^\gamma$:
\bea
X_{IJKL}&\sim &{\int  \prod_{i=1}^6 d^4 p_i \delta(p_i^2)}  \; \delta(s_{12}+s_{13} +s_{23}) \,  \delta(s_{65}+s_{64} +s_{54}) \, \delta^4 \Big (\sum_{i=1}^6 p_i \Big ) \nonumber\\
&& \bar h
(p_1)  \bar h(p_6) A_{IJ}(p_2) A_{KL}(p_5) h(p_3) h(p_4)
 [16]^2 [12]^2 [65]^2 \langle 45\rangle^4  \langle 23\rangle^4
\eea
The conjugate term, the second one in (\ref{int}) will depend on $\bar A$ and will not contribute to derivative over $A(p)$. Therefore there is only one term in this sector of the NGZ identity and it does not vanish unless $x_3=0$.

We may  interpret the expressions above as follows: we compute the value of the operator
$X_{IJKL}(0)$ between physical states of 4 gravitons and two vectors $A_{IJ}(q_2)$ and $ A_{KL}(q_5)$
\be \boxed{
\langle 0 | \partial _\mu J^\mu (0)|p_1, p_2, p_3, p_4, p_5, p_6 \rangle = x_3^2 \,  \kappa^8 \, [16]^2 [12]^2 [65]^2 \langle 45\rangle^4  \langle 23\rangle^4 }
\label{matrix3}\ee
$\delta^4 \Big (\sum_{i=1}^6 p_i \Big ) $ is a consequence of the fact that the operator $ \partial _\mu J^\mu(0)$ does not carry away any momenta. For all 6 particles $\delta(p_i^2)$ is a condition of the physical state and
$s_{12}+s_{13} +s_{23}=0$ and $s_{65}+s_{64} +s_{54}=0$
restrictions for each of the 3 particles follow from the fact that the vector $F(p)$ in the procedure of derivation of the dual field strength $G$ had null momenta. The matrix element of this operator does not vanish and therefore the $E_{7(7)}$ current conservation identity in the form (\ref{int}) is not satisfied unless $x_3=0$.

\section{L-loop 4-point case}
\subsection{Higher order polynomials of Mandelstam variables}
Green and Vanhove have established the structure of the 4-point type II superstring tree amplitude in \cite{Green:1999pv}. We will use the corresponding relation for the analysis of the N=8 supergravity counterterms. In terms of Mandelstam variables we define two linearly independent dimensionless functionals  of $s,t,u$
\be
I_2= (\kappa^2)^2 (s^2+t^2+u^2)\, \qquad I_3= (\kappa^{2})^3 (s^3+t^3+u^3)
\ee
where
\be
s+t+u =0\,  ,  \qquad s^3+t^3+u^3 = 3\, stu
\ee
Any higher order polynomial $I_k= \kappa^{2k}(s^k+t^k+u^k)$ can be expressed as a functional of powers of $I_2$ and $I_3$
\be
I_k\equiv \kappa^{2k}(s^k+t^k+u^k)=  k \sum_{2p+3q=k}{(p+q-1)!\over p! \, q!} \Big ({I_2\over 2}\Big )^p \Big ({I_3\over 3}\Big )^q
\label{GVH}\ee
 At the $L$-loop level we need the insertion of $I_k$  into the 3-loop counterterm $R^4+...$,  where $k=L-3$.

The number of kinematical structures $l$ appearing at each order $D^{2k} (R^4+...)$  is given by the number of ways $k$ decomposes as the sum of a multiple of 2 and a multiple of 3, $k = 2p+3q$ (so that $I_2^p I_3^q$  corresponds to the order $s^kR^4$.

This means that there is the following pattern
\bea
L=7&& \qquad  p=2\, \qquad q=0\, , \qquad (I_2)^2\nonumber \\
L=8&& \qquad  p=1\, \qquad q=1\, , \qquad I_2 \, I_3\nonumber \\
L=9&& \qquad  p=3\, \qquad q=0\, , \qquad (I_2)^3\, ;  \qquad  p=0\, \qquad q=2\, , \qquad (I_3)^2
\nonumber \\
L=10&& \qquad  p=2\, \qquad q=1\, , \qquad I_2^2 \, I_3\nonumber \\
L=11&& \qquad  p=4\, \qquad q=0\, , \qquad I_2^4 \, ;  \, \; \qquad  p=1\, \qquad q=2\, , \qquad I_2(I_3)^2 \qquad \rm {etc}
\eea

\subsection{7-loop}
The linearized 4-point counterterm differs from the 3-loop one by the insertion of the $(s^4+t^4+u^4)$
polynomial, see eq. (4.28) in \cite{Kallosh:2009db}. In 2 graviton-2-vector sector this means 
\be
S^{7-loop}_{ _{\partial^{8}(\partial F)^2 R^2}} = x_7 \, \kappa^{12}  \int \prod_{i=1}^{4} d^4p_i \delta \Big (\sum p_i \Big ) R(p_1) \partial F(p_2) \bar R(p_3) \partial\bar F(p_4)  (s^4+t^4+u^4)\,.
\ee
Here we have one insertion in the counterterm, $(I_2)^2$. Therefore the contribution of the 7-loop CT to NGZ identity is
\bea
 && x_7^2 \, \kappa^{8}\, {\int  \prod_{i=1}^6 d^4 p_i \delta(p_i^2)}  \; \delta(s_{12}+s_{13} +s_{23}) \,  \delta(s_{65}+s_{64} +s_{54}) \, \delta^4 \Big (\sum_{i=1}^6 p_i \Big ) \nonumber\\
&& \bar h
(p_1)  \bar h(p_6) A_{IJ}(p_2) A_{KL}(p_5) h(p_3) h(p_4)
 [16]^2 [12]^2 [65]^2 \langle 45\rangle^4  \langle 23\rangle^4   B^{IJKL} (I_2^{123})^2 (I_2^{456})^2
\eea
where
\be
I_2^{123}= \kappa^4(s_{12}^2+s_{13}^2 +s_{23}^2)\, , \qquad I_2^{654}=\kappa^4( s_{65}^2+s_{64}^2 +s_{54}^2)
\ee
 There is only one term which does not  vanish, therefore it is required that $x_7=0$.

 The relation to the current conservation is, as before
 \be
\langle 0 | \partial _\mu J^\mu (0)|p_1, p_2, p_3, p_4, p_5, p_6 \rangle = x_7^2  \, \kappa^{8}\,  [16]^2 [12]^2 [65]^2 \langle 45\rangle^4  \langle 23\rangle^4  (I_2^{123})^2 (I_2^{456})^2
\label{matrix}\ee
and the restrictions $s_{12}+s_{13} +s_{23}=0$ and  $s_{65}+s_{64} +s_{54}=0$ apply.
The matrix element of this operator does not vanish and therefore the $E_{7(7)}$ current conservation identity  is not satisfied unless $x_7=0$.

\subsection{8-loop }
Here we have one insertion in the counterterm, a product, $I_2 \, I_3$.  In the identity we have to insert one combination
\be
I_2^{123} I_3^{123} I_2^{654} I_3^{654}
\ee
As in the previous case we have  a unique insertion. There is only one term which does not  vanish, therefore it is required that $x_8=0$.
\subsection{9-loop }
The insertion into the 4-point counterterm has two linearly independent structures
\be
(I_2)^3 +\alpha (I_3)^2
\ee
where $\alpha$ is an arbitrary constant parameter. In the center of mass system these two expressions correspond to the independent polynomials in $\cos\theta$.
In the identity the insertion is
\be
\Big [(I_2^{123})^3 +\alpha (I_3^{123})^2\Big ] \Big [ (I_2^{654})^3 +\alpha (I_3^{654})^2\Big ]
\ee
Can we make a choice of one constant $\alpha$ which will make the contribution to the identity of the 9-loop counterterm vanishing? We have a quadratic equation:
\be
\alpha ^2 (I_3^{123})^2 (I_3^{654})^2 +\alpha \Big ( (I_3^{123})^2(I_2^{654})^3 + (I_3^{654})^2 (I_2^{123})^3\Big ) +
(I_2^{123})^3 I_2^{654})^3=0
\ee
The solution of this equation for $\alpha$ is a functional of momenta,  there is no solution for constant $\alpha$.

\subsection{L-loop}

In the L-loop 4-point case we will have some finite number of constants $\alpha_1, ..., \alpha_l$. 
\be
S^{L-loop}_{ _{\partial^{2(L-3)}(\partial F)^2 R^2}} = x_L \, \kappa^4 \int \prod_{i=1}^{4} d^4p_i \delta \Big (\sum p_i \Big ) R(p_1) \partial F(p_2) \bar R(p_3) \partial\bar F(p_4)  (J_0 +\alpha_1 J_1 +...\alpha_l J_l\Big )\,.
\label{L}\ee
Here $J_i$ are dimensionless polynomials of the type $\kappa^{2(L-3)}p^{2(L-3)}$ and $l$ is the number of independent ones, according to (\ref{GVH}). The dual field strength is
\be
\hat{G}(p) = x_L \, \kappa^4  \int\prod_{i=1}^{3} d^4p_i \delta(p+ \sum_{i=1}^{3} p_i)R(p_1) \partial \partial F(p_2) \bar R(p_3)  (J_0 +\alpha_1 J_1 +...\alpha_l J_l\Big ).
\label{Dan1}\ee
This provides the left hand side of the  current conservation identity
 \be
\langle 0 | \partial _\mu J^\mu (0)|p_1, p_2, p_3, p_4, p_5, p_6 \rangle = x_L^2  \, \kappa^{8}\,  [16]^2 [12]^2 [65]^2 \langle 45\rangle^4  \langle 23\rangle^4  \Big (J_0 +\alpha_1 J_1 +...\alpha_l J_l\Big ) \Big (K_0 +\alpha_1 K_1 +...\alpha_l K_l\Big )
\label{matrixL}\ee
where $J_i, \, i=0,...,l$ depend on $s_{12}, s_{13}, s_{23}$ and $K_i, \, i=0,...,l$ depend on $s_{65}, s_{64}, s_{54}$
and the restrictions $s_{12}+s_{13} +s_{23}=0$ and  $s_{65}+s_{64} +s_{54}=0$ apply.
For this to vanish we have to require that for all $p_1,...,p_6$ consistent with constraints, the following is true
\be
\Big (J_0 +\alpha_1 J_1 +...\alpha_l J_l\Big ) \Big (K_0 +\alpha_1 K_1 +...\alpha_l K_l\Big )=0
 \label{zero}\ee
The constraints require each of the 6 momenta are null and both combination of 3 momenta  have a vanishing $s+t+u$.
These constraints are not so strong  as to require all $J_i$ and $K_i$ to be constants. Therefore eq. (\ref{zero}) has no solutions for any finite number $l$ of constants $\alpha_i$. The current conservation identity is violated unless $x_L$ for the $L$-loop 4-point counterterm vanishes.

\section{Non-linear versus linear  counterterms}

Generic non-linear L-loop counterterms  \cite{Howe:1980th,{Kallosh:1980fi}} in the on-shell superspace \cite{Brink:1979nt}  have the following form
\be
\kappa^{2(L-1)}  \int d^4x \, d^{32} \theta \,  {\rm Ber} E \, {\cal L}^{\rm CT}_L  \Big (T^P_{KL}(x, \theta) , R_{PQKL} (x, \theta)\Big ) \ .
\label{CT}\ee
Here ${\rm Ber} \, E$ is the super-determinant and $T^P_{KL}(x, \theta)$ and $R_{PQKL} (x, \theta)$ are the components of the superspace torsion and superspace curvature, respectively. The lowest dimension superfield describing the superspace torsion  starts with gaugino $\chi^{\alpha}_{ijk}$ and has dimension ${1\over 2}$. The Bianchi identities have been solved in the superspace \cite {Brink:1979nt} and they turned out to be equivalent to non-linear classical equations of motion for superfields.
Superspace torsion and curvature live in the tangent space: this means that they transform under Lorentz transformations and under $SU(8)$ transformations. Each torsion and curvature is invariant under curved superspace transformations: general covariance, non-linear local supersymmetry and $E_{7(7)}$ symmetry.

All geometric torsions and curvatures in superspace are non-linear functions of the following superfields:
fermion  superfields of dimension 1/2 ,  $\chi_{\alpha ijk}, \; \bar \chi_{\dot \alpha}^{ijk}$ ,  and of dimension 3/2 , $\psi_{\alpha\beta\gamma i }, \; \bar \psi_{\dot \alpha \dot \beta \dot \gamma} ^i $,  and their derivatives $D_{\alpha \dot \beta}$.  The   bosonic superfields  have dimension 1, $F_{\alpha\beta \, ij} , \; \bar F_{\dot \alpha\dot \beta}^{ij}$ and  $P_{\alpha \dot \beta ijkl},  \; \bar P_{\alpha \dot \beta}{}^{ ijkl} $  and dimension 2 , $R_{\alpha \beta \gamma\delta}, \; \bar R_{\dot \alpha \dot \beta \dot \gamma \dot \delta} $, and their derivatives.
These are our building blocks.

The mass dimension of the L-loop counterterm ${\cal L}_{\rm CT}$ in superspace is computes as follows: $[\kappa^{2(L-1)}]= -2(L-1), \; [d^4x]= -4,  [d^{32}\theta ]=+16$ so that
\be
[{\cal L}^{\rm CT}_L(x, \theta)]=   2(L-1)-16+4= 2 (L-7)
\label{dim}\ee
 It is important that the scalar superfield $W_{ijkl}(x, \theta)= \phi_{ijkl}(x)+...$ which starts with the scalar superfield and  is dimensionless, does not belong to the superspace geometry. Effectively this means that only derivatives of the scalars enter in linearized form of the non-linear counterterms in agreement with linearized $E_{7(7)}$ symmetry. At the non-linear level only the superfield $P_{\alpha \dot \beta ijkl}(x, \theta)$  and its conjugate, which are both $SU(8)$ tensors and  $E_{7(7)}$ invariants, may appear in the counterterms. It is worth reminding here that the scalar part of the classical action is given by
 \be
 {\cal L}_{cl}^{scalar}(x) = P_{\alpha \dot \beta ijkl} \bar P^{\alpha \dot \beta ijkl}(x)
 \ee
where $P_{\alpha \dot \beta ijkl}(x)= \partial_{\alpha \dot \beta } \phi_{ijkl}(x)+...$ is an infinite power series in scalars forming the coset space ${E_{7(7)}\over SU(8)}$. Using the vielbein structure of ${E_{7(7)}\over SU(8)}$
\be
 {\cal L}_{cl}^{scalar}(x)= Tr \Big (D_\mu V V^{-1} D^\mu V V^{-1} \Big )
\ee
where the vielbein $V$ transforms under the local $SU(8)$ from the left and under and global $E_{7(7)}$ from the right
\be
V'= U(x) V E^{-1}
\ee
Under the linearized $E_{7(7)}$ symmetry the dimensional scalar field transforms
\be
\delta_{E_{7(7)}} \phi_{ijkl}(x)= \Sigma_{ijkl}\, ,  \qquad [\phi_{ijkl}]=0
\ee
Under the exact $E_{7(7)}$ symmetry the geometric superfields are invariant, in particular
\be
 (\chi^{\alpha}_{ijk})' = \chi^{\alpha}_{ijk} \,  ,\qquad  [\chi^{\alpha}_{ijk}]=1/2  \,  ,\qquad   P_{\alpha \dot \beta ijkl} '= P_{\alpha \dot \beta ijkl}\,  ,\qquad [P_{\alpha \dot \beta ijkl}]=1
  \,  ,\qquad \rm etc
\ee
{\it They all have positive dimension which explains why the number of non-linear invariants at every loop level is limited, whereas the number of linearized invariants at a given loop level is infinite.}

It may be instructive to present here  the non-linear local supersymmetry transformation of gaugini which is
\be
\delta_{susy} \chi_{\alpha\,  ijk}(x)  = \epsilon_{[i} {}^\beta F_{\alpha \beta jk]}(x) +\bar \epsilon{}^{\dot \beta l} P_{\alpha \dot \beta ijkl}(x)
\label{nonlin}\ee
As explained above, $P_{\alpha \dot \beta ijkl}(x)= \partial_{\alpha \dot \beta } \phi_{ijkl}(x)+...$ is an infinite power series in scalars forming the coset space ${E_{7(7)}\over SU(8)}$ since $P$ is a part of an $SU(8)$ covariant 1-form $(DV) V^{-1}$. In the linear approximation only the first term of the 1-form $P$ remains which is a basis for linearized supersymmetry of the amplitudes. But there is an infinite number of terms in the non-linear supersymmetry (\ref{nonlin}) which is taken care by geometric methods. In particular, in the superspace the fermionic derivatives of the gaugini superfield are
\be
D_{\beta }^l \chi_{\alpha\,  ijk}(x, \theta)= \delta ^l{}_{[i} F_{\alpha \beta  kl]} (x, \theta)\, , \qquad D_{\dot \beta l } \chi_{\alpha\,  ijk}(x, \theta)= P_{\alpha \dot \beta ijkl}(x, \theta)
\ee
Using the recent advances in computing amplitudes in the helicity formalism the linearized form of all these counterterms was constructed and analyzed recently in \cite{Elvang:2010jv}. The purpose here is to compare the information obtained in both methods, in \cite{Howe:1980th,Kallosh:1980fi,Kallosh:2010kk}  and in \cite{Elvang:2010jv}.

In both cases, in superspace as well as in amplitude structures one can clearly see the proliferation of counterterms with increasing number of loops and legs. A simple explanation of this proliferation of the candidate UV divergences is: in superspace  with more loops and more legs  more scalars  can be build from torsions of curvatures.

\subsection{$L=7, n>4$}

The case $L=7$ is special. The Lagrangian must have dimension zero since
\be
[{\cal L}^{\rm CT}_7]= 2 (L-7)=  0
\ee
It cannot depend on torsions and/or curvatures, which always have positive dimension. Therefore
 the full non-linear structure of the counterterm is unique and given by the superspace volume
\be
S_7^{CT} =x_7 \, \kappa^{12} \int d^4x \, d^{32} \theta \,  {\rm Ber} E
\label{volume}\ee
It is not clear if it exists\footnote{In view of our current analysis of  $E_{7(7)}$ current conservation it does not matter, this candidate counterterm will not support the 7-loop UV divergence.}
since for N=2 supergravity the volume of the superspace $  \int d^4x \, d^{8} \theta \,  {\rm Ber} E$ has been proven to vanish \cite{Sokatchev:1980td}.
We may now expand this expression to present the linearized form of each $n$-point amplitude.
\be
S_7^{CT}= x_7\,  \kappa^{12} \int d^4x \, d^{32} \theta  [  W_{ijkl}^4 +  W_{ijkl}^5 + ... W_{ijkl}^n+...]
\label{7}\ee
Here $W_{ijkl}^n$ is the symbolic expression for the $SU(8)$ invariants constructed from $n$ superfields
$W_{ijkl}$. No derivatives in $x$ or $\theta$ direction are permitted since they will raise the dimension of $[L^{\rm CT}_7]$ which has to vanish.

If, however, we are interested only in linear supersymmetry, we may consider the following infinite number of linearized superinvariants
\be
S_7^{CTlin}=   \kappa^{12} \int d^4x \, d^{32} \theta  [ a_4 W_{ijkl}^4 + a_5 W_{ijkl}^5 + ... a_nW_{ijkl}^n+...]
\label{7l}\ee
The coefficients $a_n$ are arbitrary and independent since each term in (\ref{7l}) has only linear supersymmetry, whereas (\ref{volume}) has a non-linear one.

Let us compare the properties of the linearized counterterms with $n$-points which follow from (\ref{7l}) with those discussed in \cite{Elvang:2010jv}. For example, for $n>8$ there are no pure graviton amplitudes as one can see
by dimensional reason since
$S= \kappa^{12} \int d^4 x R^{n}  $ for $n>8$ has positive dimension. In superspace the scalar superfield depends on curvature as follows
\be
W\sim \theta^4 R+ \theta^6 \partial R+...\, ,  \qquad W^n \sim (\theta^4 R + \theta^6 \partial R+...)^n
\label{dic}\ee
 For $n>8$ we have $\theta^{4n} R^n+...$,  which vanishes at $n>8$.

The 4-point 7-loop CT was argued to be absent to comply with the $E_{7(7)}$ NGZ identity \cite{Kallosh:2011dp}. Since the $n$-point amplitudes with $n>4$ are defined by the same unique expression in (\ref{7}), it means that all $n>4$ UV divergences at 7-loop must be absent.

Another way to see the same is to apply the argument in \cite{Kallosh:2010kk}
 and in \cite{Elvang:2010jv} that all $L=7$, $n>4$ linearized CT's violate linearized $E_{7(7)}$. Either way, we conclude that 7-loop level is predicted to be clear from UV divergences according to $E_{7(7)}$ current conservation.

\subsection{$L=8, n>4$}
The counterterm which at the linear level has 4-point terms is given by (\ref{CT}) where
  \cite{Howe:1980th,Kallosh:1980fi}
\be
{\cal L}^{\rm CT}_8= x_8 \,   \chi_{ijk}^\alpha \, \chi_{\alpha mnl} \, \bar \chi^{\dot \alpha pqr }
\bar \chi_{\dot \alpha}^ {stu } X^{ijkmnl}{}_{pqrstu} \qquad [{\cal L}^{\rm CT}_8]=   2(8-7)=2
\label{chi4}\ee
 It is a product of 4 super-torsions (each has dimension 1/2) and  $X^{ijkmnl}{}_{pqrstu}$ is some numerical $SU(8)$ tensor which provides an $SU(8)$ invariant combination of the 4 $SU(8)$ gaugini's. This counterterm is non-linear, it  has an infinite
amount of higher point terms, $n$-point amplitudes, which complete the linearized 4-point expression. But all of them come with the same $x_8$ as the 4-point part of it.
There are no other non-linear invariants at the 8 loop level since adding derivatives or more torsions/curvatures will raise dimension above 2.

So, what is the relation between the statement above that at the  8-loop  there is a unique 4-point operator whose nonlinear corrections produce all $n > 4$ point matrix elements, and
 statement in \cite{Elvang:2010jv} that there are many $n>4$ linearized invariants?

 The crucial point here is that for linear level supersymmetric invariants we may use a {\it dimensionless superfield $W_{ijkl}(x, \theta)$. Any power of $W$ has dimension zero!} Therefore we may have many linearized invariants of the form
 \be
S_8^{CTlin}= \,  \kappa^{14} \int d^4x \, d^{32} \theta   D^2
 [ a_4 W_{ijkl}^4  + a_5 W_{ijkl}^5 + ... a_nW_{ijkl}^n+...] \label{linear8} \ee
Here $D^2$ can be replaced by 4 spinorial derivatives, $D_\alpha^2\bar D_{\dot \alpha}^2$ and one should keep in mind that the derivatives may act at any of the superfields. Each $a_n$ is independent since we only require linear supersymmetry from each CT.

However, {\it to build the non-linear invariant we are allowed to use only geometric superfields with the minimal dimension 1/2}. This allows according to (\ref{chi4}) only 4 gaugini's. Any other combination of $n>4$ geometric superfields will have dimension $>2$ in contradiction with $[{\cal L}^{\rm CT}_8]=   2(8-7)=2$.

It is interesting to see some cases, for example, using (\ref{dic})
 \be
S_8^{CTlin}= \,  \kappa^{14} a_6  \int d^4x \, d^{32} \theta   D^2
   W_{ijkl}^6    \sim  \kappa^{14} a_6  \int d^4x \, D^6 R^6+...
\ee
 \be
S_8^{CTlin}= \,  \kappa^{14} a_8  \int d^4x \, d^{32} \theta   D^2
   W_{ijkl}^8    \sim  \kappa^{14} a_8  \int d^4x \, D^2 R^8+...
\ee
We may also give here examples from the rhs part of the Table I in the second reference in \cite{Elvang:2010jv}
 \be
S_7^{CTlin}= \,  \kappa^{12} a_{16}  \int d^4x \, d^{32} \theta
   W_{ijkl}^{16}    \sim  \kappa^{12} a_{16 } \int d^4x \, \phi^8 R^8+...
\ee
and
 \be
S_8^{CTlin}= \,  \kappa^{14} a_{14}  \int d^4x \, d^{32} \theta
   W_{ijkl}^{14}    \sim  \kappa^{14} a_{14 } \int d^4x \, \phi^6 R^8+...
\ee
The non-linear supersymmetry does not allow to use the dimensionless $W$, only geometric ones with dimension $\geq 1/2$ are allowed. This explains the uniqueness of the non-linear L=8 CT  in (\ref{chi4}) and  that there are many $n>4$ linearized invariants in \cite{Elvang:2010jv} as well as in the linearized superspace as shown in (\ref{linear8}).

 We have explained the relation between the linearized counterterms in Table I in \cite{Elvang:2010jv} and superspace ones, including the entries on the extreme  right hand side of this Table.
The major difference between the linearized counterterms in Table I in  \cite{Elvang:2010jv} and the complete non-linear ones in   \cite{Howe:1980th,Kallosh:1980fi} is the level of supersymmetry: in  \cite{Elvang:2010jv} the supersymmetry is global, it relates at the linear level the amplitudes of particles which belong to the same supermultiplet. The supersymmetry in \cite{Howe:1980th,Kallosh:1980fi} is local non-linear, it is much more restricted, since it requires an infinite number of terms for completion. This is achieved in superspace which is controlled by the non-linear ${E_{7(7)}\over SU(8)}$ coset space geometry. In particular, it rules out all linear superinvariants, as counterterms, for $L=7,8$, as we have shown above.

\section{Discussion}

In this paper we gave important explicit examples of the general Lorentz covariant proof in \cite{Kallosh:2011dp} of the UV finiteness of perturbative N=8 supergravity. Whereas in  \cite{Kallosh:2011dp} we used the superfields  and coordinate space, in this paper we use the helicity formalism in the momentum space, which allows to make some details of our general approach more transparent.

In particular, we tested the $E_{7(7)}$ current conservation identity in the form (\ref{critical}) for the deformed N=8 supergravity where the candidate counterterms are added to the classical action.
Using the helicity formalism in the momentum space we computed the contribution of the 4-point $3$-loop candidate counterterms to this identity. The contribution to a dual field strength in a sector of the theory with 2-gravitons and 2 vectors is given by a simple in momentum space unique  expression in (\ref{Dan}). As the result,   the current conservation identity for the 3-loop case  acquires a simple form shown  in eq. (\ref{matrix3}). The left hand side does not vanish unless $x_3=0$ and there is no 3-loop UV divergence.

Computing the higher loop order $L$ contribution to the 4-point amplitudes requires an insertion into an $E_{7(7)}$ identity of various powers of the higher order polynomials $(s^k+ t^k + u^k)$ where $k=L-3$, see  eq. (\ref{matrixL}). The number of independent  polynomials
is given by the number of ways $L-3$ decomposes as the sum of a multiple of 2 and a multiple of 3, $L-3 = 2p+3q$,  as shown in \cite{Green:1999pv}. This number, however, is discrete,  and cannot help  to solve the equation
(\ref{zero}) which has to be valid at continuous values of the Mandelstam variables $s_{ij}$. Therefore all $L$-loop 4-point candidate counterterms in (\ref{L}) violate $E_{7(7)}$ current conservation and do not support the corresponding UV divergences.  $E_{7(7)}$ current conservation requires $x_L=0$. 

In case of $L=7,8$ where complete non-linear candidate counterterms are unique and defined by their 4-point parts \cite{Howe:1980th,Kallosh:1980fi}, the current conservation identity forbids all $n$-point candidate counterterms.

\section*{Acknowledgments}
We are grateful to  D. Freedman for his help with setting up a momentum space computation to test the $E_{7(7)}$ current conservation identity and also for suggesting to clarify the relation between linearized and complete non-linear superinvariants  as candidate counterterms. We are grateful S. Ferrara and A. Van Proeyen for the insightful questions on  the technical aspects of our related paper \cite{Kallosh:2011dp} and for their support.
 We appreciate stimulating discussions with L. Brink, M. Henneaux,  A. Linde, H. Nicolai and L. Susskind.  This work  is supported by the NSF grant 0756174.





\end{document}